\documentclass{article}
\usepackage{amssymb}
\usepackage{amsfonts}
\usepackage{amsmath}

\begin{document}

\title{\textbf{The beryllium hollow-body solar sail: exploration of the
Sun's gravitational focus and the inner Oort Cloud}}
\author{\textbf{Gregory L. Matloff} and \textbf{Roman Ya. Kezerashvili} \\
{\small Physics Department, New York City College of Technology,}\\
{\small \ The City University of New York}\\
{\small \ 300 Jay Street, Brooklyn, NY 11201}\\
{\small GMatloff@CityTech.Cuny.Edu, RKezerashvili@citytech.cuny.edu} \and 
\textbf{Claudio Maccone} \\
{\small Member of the International Academy of Astronautics,}\\
{\small \ Via Martorelli 43, Torino (TO), 10155, Italy }\\
{\small \ clmaccon@libero.it} \and \textbf{Les Johnson} \\
{\small NASA Marshall Space Flight Center, Huntsville, AL 35812, USA}\\
{\small C.Les.Johnson@nasa.gov}}
\maketitle

\begin{abstract}
Spacecraft kinematics, peak perihelion temperature and space environment
effects during solar-radiation-pressure acceleration for a beryllium
hollow-body interstellar solar sail inflated with hydrogen fill gas are
investigated. We demonstrate that diffusion is alleviated by an on-board
fill gas reserve and electrostatic pressure can be alleviated by increasing
perihelion distance. For a 0.1 AU perihelion, a 937 m radius sail with a
sail mass of 150 kg and a payload mass of 150 kg, perihelion sail
temperature is about 1000 K, peak acceleration is about 0.6 g, and
solar-system exit velocity is about 400 km/s. After sail deployments, the
craft reaches the 200 AU heliopause in 2.5 years, the Sun's inner
gravitational focus at 550 AU in about 6.5 years and 2,550 AU in 30 years.
The Be hollow-body sail could be applied in the post 2040 time frame to
verify general relativity predictions regarding the Sun's inner
gravitational focus and to explore particles and fields in the Sun's inner
Oort Comet Cloud.
\end{abstract}

\section{Introduction: an approach to extrasolar space exploration}

The solar sail has emerged as one of the few ultimately feasible modes of
extrasolar space exploration and travel. As demonstrated by Matloff and
Mallove \cite{1}, \cite{2} a parachute-like space-manufactured metallic sail
unfurled at the 0.01 AU perihelion of an initially parabolic solar orbit can
reach solar-system exit velocities of about 0.003$c$-0.004$c$. Furthermore,
sail and diamond-strength cables can be conceptually woven around payload to
serve as cosmic-ray shielding during the long interstellar transfer and
unfurled later for deceleration at a solar-type star \cite{3}. Additional
work on this concept has included studies of sail materials and spacecraft
pre-perihelion trajectories.

The analytical studies of interstellar solar sailing led to the European
ASTROsail and SETIsail extrasolar probe concepts of the early 1990's, which
would have explored the Sun's near gravitational focus at 550 AU \cite{4}.
In turn, these studies were superseded by the less ambitious European Aurora
sail, which would have explored the Sun's heliopause, the boundary between
solar and galactic influence at about 200 AU \cite{5}. Aspects of Aurora
have been incorporated in the planned NASA Heliopause Sail, which could be
launched around 2020 \cite{6}.

For true extrasolar travel, the parachute-type sail of the early studies may
not be ideal. Such craft do not scale easily between small, near term probes
and ultimate human-occupied generation ships. The proportion of spacecraft
mass that must be devoted to cable increases rapidly with payload \cite{7}.
The square-rigged NASA concept is also not ideal---it may not easily scale
from low-acceleration extrasolar probes to ultimate higher acceleration
generation ships.

\section{The hollow-body solar sail}

Strobl in Ref. \cite{8} suggested a type of inflatable interstellar solar
sail variously called the "hollow-body"\ or "pillow"\ sail. In a hollow-body
sail, the Sun-facing sail surface is reflective and the space-facing sail
surface is emissive. Both sail surfaces have thicknesses measured in tens of
nanometers. The sail is inflated by low-pressure fill gas. Most studies to
date have assumed hydrogen fill gas to minimize fill-gas mass. Payload is
mounted on the space-facing sail surface. Unlike most solar photon sails,
which pull the payload through space during acceleration, the hollow-body is
a "pusher"\ sail.

In order to minimize sail perihelion, in Ref. \cite{8} assumed that the sail
would be constructed using metals optimized for very high melting points. A
disadvantage of this approach is that most such metals have high specific
gravities, which reduces solar-system exit velocity.

Recently, in Ref. \cite{9} evaluated an interstellar hollow-body solar
photon sail constructed from beryllium. From a thermal point of view, a 0.05
AU perihelion was possible for beryllium hollow-body sails with a 20 nm
thick Sun-facing surface and a 10 nm thick space-facing surface. Two
configurations of the beryllium hollow-body solar photon sail were
considered. Configuration A, a small generation ship, had a fully inflated
sail radius of 541.5 km, a payload of $10^{7}$ kg, a separation between sail
faces of 1 km and 30,000 kg of molecular hydrogen fill gas. Configuration B,
a near-term extrasolar probe, had a sail radius of 937 m, a 30 kg payload, a
1.8 m separation between sail faces and 0.16 grams of hydrogen fill gas.

From the point of view of kinematics, mechanical stress, and thermal
effects, the hollow-body solar photon sail scales well. Both configurations
had a spacecraft areal mass density of $6.52\times 10^{-5}$ kg/m$^{2}$, a
peak internal gas pressure of $1.98\times 10^{-4}$ Pa, and a peak perihelion
temperature of 1412 K. If fully inflated at the 0.05 AU perihelion of an
initially parabolic solar orbit, both had a peak radiation-pressure
acceleration of 36.4 m/s$^{2}$ and exited the solar system at 0.00264$c$
after an acceleration duration less than one day.

\section{The hollow-body sail and the space environment}

In a series of recent papers \cite{10}, \cite{11} dynamics and
space-environment effects of a beryllium hollow-body interstellar solar sail
inflated with hydrogen fill gas have been investigated for the case of sail
deployment at the 0.05 AU perihelion of an initially parabolic solar orbit
and the interaction of the solar radiation with the solar sail materials and
the hydrogen fill gas was studied. These analyses evaluate worst-case solar
radiation effects during solar-radiation-pressure acceleration. The
diversity of physical processes of the interaction of photons, electrons and
protons with the sail leading to electric charging of the sail material are
analyzed. Issues include diffusion of hydrogen fill gas through the 10-20 nm
sail walls at elevated perihelion temperatures and electrostatic pressure
from sail charge build-up, which is an issue because beryllium's tensile
strength decreases with increasing temperature. It was realized that also
necessary to analyze the interaction between the hollow-body sail and the
near-perihelion space environment \cite{10}, \cite{11}. It was assumed in
these studies that near-Sun missions would most likely be conducted during
Quiet-Sun periods. The near-Sun environment, even under Quiet-Sun
conditions, is a most dynamic region of space. Spacecraft designers must
cope with both the solar wind and copious amounts of ionizing solar
radiation. In Refs. \cite{10}, \cite{11}, the perihelion pass was modeled
using the simplifying, but conservative assumption of 7,000 seconds at a
constant perihelion distance. The solar wind at 1 AU was assumed to have a
velocity of 400 km/s relative to the Sun and an ion density of 10 ions per
cubic centimeter. Solar wind velocity was assumed to be constant with solar
distance, and ion density was assumed to follow an inverse square law.

Interactions between the sail and ionizing solar radiation were analyzed, as
were interactions between solar wind ions and the sail. It was also
necessary to consider the interactions of the solar radiation and solar wind
with the hydrogen fill gas within the sail. It was found that the
hollow-body solar photon sail does not scale well when interactions between
the hydrogen fill gas and solar radiation and solar wind are considered.
Diffusion of hydrogen through sail walls at perihelion has little effect
upon the large Configuration A. But, in the worst case, all hydrogen within
the small Configuration B sail diffuses through the sail walls within about
70 seconds at a 0.05 AU perihelion. This problem could be alleviated by
carrying a large hydrogen reserve. Both configurations are also limited by
electrostatic pressure, which bursts the sail in the worst case. Although
other mitigation approaches are possible, one approach to alleviating this
problem is to increase perihelion distance, at the expense of performance.

\section{The Oort Cloud explorer - gravity focus probe}

We consider a modified Configuration B beryllium hollow-body solar-photon
sail with a sail radius of 937 m and a sail mass of 150 kg. To perform a
greater range of scientific observations, the payload mass is increased to
150 kg. Therefore, the total spacecraft mass is 300 kg. The perihelion has
been increased to 0.1 AU. As is shown below, this action greatly alleviates
the electrostatic pressure issue.

\subsection{Kinematics}

First we consider spacecraft kinematics by defining the lightness number $%
\eta _{sail}$, which is the ratio of solar radiation pressure force on the
sail to solar gravitational force on the spacecraft, assuming that the
(opaque) sail is normal to the Sun \cite{12}, \cite{7} and \cite{9}:

\begin{equation}
{\eta }_{sail}{=}\frac{(1+R_{sail})}{c\sigma _{sail}GM_{Sun}}SR^{2},
\end{equation}%
where $R_{sail}$ is sail reflectivity to sunlight, $c$ is the speed of
light, $\sigma _{sail}$ is the spacecraft areal mass thickness, $G$ is the
universal gravitational constant, $M_{Sun}$ is the solar mass, $S$ is the
average value of the Poynting vector, which is the average power per unit
area transported by the solar light, and $R$ is the distance between the
spacecraft and the Sun's center. At $R_{0}=1$ AU from the Sun, the average
value of the Poynting vector (the Solar constant) is approximately equal to $%
S_{0}=1,400$ watts/m$^{2}$. Applying the inverse square law we can determine
the incident solar energy flux at the distance $R$ from the Sun to the
spacecraft as

\begin{equation}
{S=}\frac{S_{0}R_{0}^{2}}{R^{2}}.
\end{equation}

Substituting Eq. (2) into Eq. (1), we obtain:

\begin{equation}
{\eta }_{sail}{=}\frac{(1+R_{sail})}{c\sigma _{sail}GM_{Sun}}S_{0}R_{0}^{2}.
\end{equation}

We next define a reflectivity factor $R_{f}$. For opaque sails, $R_{f}$ $%
=(1+R_{sail})/2$. For thinner partially transparent sails such as the one
under consideration here, $R_{f}=(A_{sail}+2R_{sail})/2$, where $A_{sail}$
is the fractional absorption of sunlight by the sail \cite{9}. Taking into
account this fact and substituting the numerical value of all constants into
Eq. (3), finally for the the lightness number we are getting:

\begin{equation}
{\eta }_{sail}\approx {1.574\times 10}^{-3}\frac{R_{f}}{\sigma _{sail}}.
\end{equation}
Assuming a fully inflated disc-shaped sail with the radius of 937 m and sail
mass of 150 kg., sail area is $2.76\times 10^{6}$ m$^{2}$. The spacecraft
areal mass density $\sigma _{sail}$ is, therefore, $1.09\times 10^{-4}$ kg/m$%
^{2}$. From \cite{9}, the reflectivity factor of the sail, $R_{f}$, is
0.636. Substituting these values of $\sigma _{s}$ and $R_{f}$ \ in Eq. (4),
we find that $\eta =9.18$. At the 0.1 AU perihelion, the solar-gravitational
acceleration on the spacecraft is 0.59 m/s$^{2}$. The
solar-radiation-pressure acceleration on the sail is therefore 5.42 m/s$^{2}$
or about $0.55g$.

Next we estimate solar-system exit velocity or interstellar cruise velocity, 
$v_{in}$. Although this can be done for a spacecraft departing the solar
system from an initially elliptical solar orbit \cite{12}, we choose here
the higher performance case of departure from an initially parabolic solar
orbit. For the case of constant sail orientation normal to the Sun during
the acceleration process we get

\begin{equation}
{v}_{in}\approx \sqrt{{\eta }_{sail}}v_{pp},
\end{equation}%
where $v_{pp}$ is the solar escape (or parabolic) velocity at perihelion.
Using the definition of escape velocity \cite{7} we can find interstellar
the cruise velocity: 
\begin{equation}
v_{in}{=4.21\times 10}^{4}\sqrt{\frac{\eta }{R_{AU}}},
\end{equation}
where $R_{AU}$ is the solar perihelion distance in Astronomical Units.
Applying this equation, our spacecraft departs the solar system at 403 km/s.
At this velocity, the spacecraft reaches the 200 AU heliopause in about 2.4
years. It passes the Sun's inner gravitational focus at 550 AU in about 6.5
years. During a thirty-year operational lifetime, the probe will travel more
than 2,500 AU and, therefore will reach the inner Oort Cloud.

\subsection{Peak perihelion temperature}

Next we consider the peak temperature at perihelion, $T_{p}$. From the
Stefan-Boltzmann law for grey bodies, sail radiant emittance is expressed as 
$W_{sail}$ $=2\varepsilon _{sail}\sigma T_{p}^{4}$ (since the sail can
radiate from both sun-facing and space-facing surfaces, there is the factor
2), where $\varepsilon _{sail}$ is the sail emissivity and $\sigma
=5.67\times 10^{-8}$ Wm$^{2}$K$^{-4}$ is the Stefan-Boltzmann constant. From
the law of conservation of energy in termal equilibrium radiant emittance
should be equil to total absorbed energy by the solar sail. Therefore, $%
W_{sail}$ $=A_{sail}S,$ where $A_{sail}$ is sail absorption which relates to
the intire sail From this condition we obtain

\begin{equation}
T{=333}\left( \frac{A_{sail}}{\varepsilon _{sail}R_{AU}^{2}}\right) ^{1/4},%
\text{ }
\end{equation}%
where $R_{AU}$ is the separation between the sail and Sun at perihelion, in
Astronomical Units. Follow \cite{9}, $A_{sail}$= 0.437 and $\varepsilon
_{sail}$= 0.530, and substituting of these values into Eq. (7) we obtain
that the temperature at the 0.1 AU perihelion is 1003 K, which is
considerably less than the 1412 K perihelion temperature previously
estimated for the 0.05 AU perihelion pass.

\subsection{Hydrogen fill gas requirement and diffusion mitigation}

The perihelion sail pressure is the ratio of radiation-pressure force on the
sail to sail area. Since the payload is 150 kg and the peak
radiation-pressure acceleration is 5.42 m/s$^{2}$, the perihelion
radiation-pressure force on the sail is 813 Newtons. Dividing this by the
area of the 937 m radius disc-shaped sail, the perihelion sail pressure is
calculated to equal $3\times 10^{-4}$ Pa, which is about 50\% higher than
the value considered by Matloff \cite{9} because of the increased payload
mass. But the number of moles of hydrogen fill gas required to maintain sail
inflation is directly proportional to sail pressure and inversely
proportional to sail temperature. Therefore, the number of moles of hydrogen
required to maintain inflation at perihelion is about 7\% or about 0.086
moles (0.17 grams) higher for a Configuration B sail than it was considered
previously.

In the worst case, as demonstrated in Ref. \cite{11}, all hydrogen fill gas
diffuses from a Configuration B sail performing a 0.05 AU perihelion pass in
about 70 seconds. Hydrogen fill gas must be replenished about 100 times
during the solar acceleration process. But as shown in Figure 2 in Ref. \cite%
{11}, the rate of hydrogen diffusion is highly dependent upon temperature
and results of calculations are very sensitive to the diffusion activation
energy and diffusion constant. For the case of a 0.1 AU perihelion, hydrogen
fill gas must be replenished about 50 times, in the worst case.

To err on the side of caution, it is assumed here that a hydrogen reserve of
100 times the required fill gas mass is carried aboard the spacecraft. This
amounts to only 170 grams of hydrogen. If hydrogen fill gas is dissociated
from water as required, no more than about one kilogram of water is
required. Even water-storage and dissociation equipment will not add more
that a few kilograms to the payload and have a very small effect on
spacecraft performance.

\subsection{Electrostatic pressure}

From a study of Kezerashvili and Matloff \cite{10} it is follows that solar
ultraviolet radiation is constantly ionizing beryllium atoms in the sail
surface during the spacecraft acceleration process and the main causes of
the ionization are photoelectric effect, Compton scattering and
electron-positron pair production processes. As a result of these processes
the surface of the solar sail will lose electrons and become charged
positively. Electrostatic pressure resulting from a net positive charge on
the beryllium sail walls may cause the sail to burst at a 0.05 AU
perihelion, because the tensile strength of beryllium degrades rapidly with
temperature. Here, we mitigate this effect by increasing perihelion distance
to 0.1 AU. As it follows from \cite{13}, the electrostatic pressure, $P$,
can be expressed as: 
\begin{equation}
P{=}\frac{\sigma _{c}^{2}}{2\varepsilon _{0}},
\end{equation}%
where $\sigma _{c}$ is the surface charge density on the sail and $%
\varepsilon _{0}$ is the electric permittivity of free space. For
Configuration A and B and a 0.05 AU perihelion, P = 59.6 MPa. which
approximates beryllium's tensile strength at 1100 K. An electron component
of the solar wind plasma partially neutralizes the positively charged sail
through radiative recombination processes as shown in Ref. \cite{11}.
Assuming that the spacecraft's pre-perihelion velocity remains parabolic,
its velocity relative to the Sun at a 0.1 AU perihelion is slower than at
the original 0.05 AU perihelion. This implies that the velocity of
solar-wind electrons at a 0.1 AU perihelion relative to the spacecraft is
greater than at a 0.05 AU perihelion. Therefore, an increase in perihelion
distance results in an increase in the flux of neutralizing electrons
striking the positive charged sail. Conservatively, this factor is ignored
here. At 0.1 AU, the flux of solar photons ionizing the sail is 0.25 times
the corresponding flux at 0.05 AU, from the inverse square law. The 0.1 AU
perihelion electrostatic pressure will therefore be less than 15 MPa, which
is much lower than the tabulated tensile strength of vacuum hard-pressed
block beryllium at 1000 K \cite{14}. Electrostatic pressure should therefore
not cause major problems at a 0.1 AU perihelion, for Quiet Sun conditions.

\section{Scientific justification}

Simply having the capability to launch a high performance extrasolar
probe---actually a prototype starship - is not enough. For such a mission to
be conducted in the foreseeable future, there must be scientific
justifications. There are at least two scientific functions this craft could
serve during its 30-year journey to 2,500 AU.

\subsection{The Sun's gravity focus}

Einstein's general relativity theory predicts that beyond 550 AU, the Sun's
gravitational field will focus and amplify emissions from distant celestial
objects occulted by the Sun. Various coronal plasma effects might push the
Sun's inner gravitational focus out as far as about 1,000 AU \cite{15}. As
well as checking relativistic predictions, this probe could perform useful
astrophysical observations if equipped with appropriate instrumentation. For
instance, many astronomers expect the existence of terrestrial planets
circling our near stellar neighbors Alpha Centauri A and B \cite{16}, \cite%
{17}. It is not impossible that solar-gravitational focusing could provide
highly resolved images of these worlds, if they exist and if the probe is
directed towards the point on the celestial sphere opposite Alpha Centauri.

\subsection{Oort Cloud exploration}

At and beyond about 1,000 AU, the probe will be within the inner fringe of
the Sun's Oort Comet Cloud. NASA has considered exploration of this
celestial region in the post-2040 time frame using a craft dubbed
``Interstellar Trailblazer.'\ Since distant Oort Cloud comets are primordial
remnants of the solar system's formation, imaging these objects in their
environment will be of interest to planetary scientists. Although Oort Cloud
comets are still gravitationally attached to our solar system, particles and
fields beyond about 200 AU will likely be of galactic rather than solar
origin. Exploration of this region could yield information regarding our
solar system's formation, evolution, and future.

\section{Conclusions}

We have demonstrated here that an appropriately configured and equipped
beryllium hollow-body solar-photon sail inflated using hydrogen fill gas
could be utilized to explore the Sun's near galactic vicinity within the
not-too-distant future. No show stoppers have been uncovered, at least from
our considerations of kinematics, thermal effects, interaction with the
space environment, and related mission aspects. There are still unknowns.
For instance, can a beryllium thin-film inflatable sail be launched from
Earth, or is in-space manufacturing necessary? Better ways may exist to
perform this mission. Is hydrogen the best fill gas for an inflatable sail
deployed near the Sun? Are there sail-wall materials superior to beryllium?
Other sail configurations than the inflatable might also be considered for
the Interstellar Trailblazer. One is a hoop sail. The hoop sail and other
contenders are worthy of further examination \cite{7}.

\end{document}